\documentclass{nsart_eng}
\usepackage{amsmath,amssymb}
\usepackage{graphicx}
\usepackage{xcolor}
\usepackage{url}

\year{2012} \volume{0} \nomer{0} \firstpage{1}

\let\leq\leqslant

\begin{document}

\title[MDI CV-QKD in optical transport network]
{Measurement-device-independent continuous variable quantum key distribution protocol operation in optical transport networks}

\author[I.~Vorontsova, R.~Goncharov, S.~Kynev, F.~Kiselev, V.~Egorov]
{$^{1}$I.~Vorontsova, $^{1}$R.~Goncharov, $^{1, 2}$S.~Kynev, $^{1, 2}$F.~Kiselev, $^{1, 2}$V.~Egorov}

\address{
$^1$ ITMO University,\\
 Kronverkskiy, 49, St. Petersburg, 197101, Russia}

\address{
$^2$ SMARTS-Quanttelecom LLC,\\
 6th Vasilyevskogo Ostrova Line, 59, St. Petersburg, 199178, Russia}

\email{iovorontsova@itmo.ru}

\udk{530.145:535.12:681.7:53.082.5}

\pacs{03.67.-a, 42.50.-p}

\begin{abstract}
Numerically, a theoretical analysis of the noise impact caused by spontaneous Raman scattering, four-wave mixing, and  linear channel crosstalk on the measurement-device-independent continuous variable quantum key distribution systems is conducted. The analysis considers symmetry and asymmetry of system paths, as well as possible channel allocation schemes, for a quantum channel located in C- and O-bans. Mathematical models for MDI CV-QKD system and the contributing noises' description are provided. The secure key generation rate is estimated to state features of protocol operation when integrated with existing DWDM systems in the context of its implementation into telecommunication networks.

\end{abstract}

\keywords{device-independence, quantum key distribution, continuous variables}

\maketitle

\section{Introduction}

Quantum key distribution (QKD) is one of the most rapidly advancing fields of quantum technologies~\cite{Scarani2009, Pirandola2020}. Its main idea is an opportunity to transport a cryptographically secure key between two and more authenticated users connected to each other through quantum and information channels. Guarantied by the principles of quantum mechanics~\cite{Gisin2002}, the security of QKD to attacks from an eavesdropper ensures safety of the transmitted data from all kinds of hacking and known attacks.

One option to classify QKD protocols is based on them being discrete-variable (DV) or continuous-variable (CV)~\cite{Grosshans2003}. Additionally, among many other QKD protocol classifications, there is the one distinguishing between protocols in terms of their device-dependence or (semi-)device-independence~\cite{Lo2012}. The intersection of these two criteria gives rise to a new class of protocols, that is measurement-device-independent (MDI) CV-QKD protocol. Device-independence features particular practical importance, for it eliminates many side channel attacks, though implies accurate theoretical analysis.

Not only does this work discuss the latter, but also for the first time combines this analysis with the task of simultaneous propagation of information and quantum signals in a single optical fiber~\cite{Mlejnek2017}. The effects considered as channel loss sources are spontaneous Raman scattering, four-wave mixing, and linear channel crosstalk. A possible realization scheme is discussed, as well as the allocation of classical channels on the standard DWDM grid. The security analysis is carried out numerically, employing the known theoretical security bounds to estimate the performance of the addressed QKD system. The results are quite important in practice to be considered when integrating QKD with the existing telecommunication networks.

\section{Measurement-device-independent QKD}

Let us discuss main principles underlying MDI QKD protocol operation~\cite{Lo2012, Ma2012}. The essence of the approach lies in the fact that no assumptions are made about the detectors involved in the protocol, such that they can even be controlled by an eavesdropper (Eve). In a typical single-photon MDI QKD protocol, two legitimate users (Alice and Bob) send quantum signals to an untrusted central relay, often addressed as Charlie. A Bell state measurement is performed then; both signals interfere at a 50:50 beam splitter (BS). Next, output signals go through a polarizing beam splitter (PBS) to be projected into either horizontal (H) or vertical (V) polarization state. The measurement is pronounced successful if two of the four involved detectors click.

\subsection{Continuous-variable MDI QKD protocol}

Similarly to the conventional MDI QKD, the continuous variable (CV) version of the protocol~\cite{Papanastasiou2017,lupo2018} again implies there are the two senders and an untrusted relay performing the measurements to be used during legitimate users' post-processing to generate the secure key. 

The two known approaches to a general protocol description, namely, “prepare-and-measure” (PM) and “entanglement-based” (EB) scenarios, apply to the case of MDI CV-QKD as well. For these scenarios are equivalent in terms of their mathematical description and effectiveness, we will consider a more practically convenient PM version of the protocol. Gaussian modulation~\cite{Grosshans2003,Weedbrook2004a} (GG02 protocol) is considered, so Alice and Bob are operating with coherent states with a two-dimensional Gaussian distribution. 
They first generate coherent states $|x_{\mathrm{A}}+ip_{\mathrm{A}}\rangle$ and $|x_{\mathrm{B}}+ip_{\mathrm{B}}\rangle$ with the quadratures $x$ and $p$ featuring variance $V_{\mathrm{A} (\mathrm{B})}-1$ (in shot noise units (SNU)) and then send their states to Charlie via quantum channels. Next, Alice's and Bob's modes interfere at the beam splitter, while Charlie measures the $\mathrm{C}$ and $\mathrm{D}$ modes' quadratures on homodyne detectors and announces the resulting state $\{X_{\mathrm{C}},\:P_{\mathrm{D}}\}$ publicly. It is Bob only who changes his state according to $X_{\mathrm{B}}=x_{\mathrm{B}}+kX_C, P_{\mathrm{B}}=p_{\mathrm{B}}-kP_{\mathrm{D}}$ (with $k$ standing for the gain associated with channel losses), whereas Alice's state remains unchanged. Finally, standard procedures are utilized for parameter estimation, information reconciliation, and privacy amplification. 

Since there is equivalency between the CV-QKD EB and PM scenarios' security proof against collective attacks~\cite{Grosshans2003a,Laudenbach2017}, we shall now switch to the well-known covariance matrix formalism.

Implying that Eve has access to the relays, quantum channels, and even Bob's state displacement operation in the EB scheme, further security analysis of the MDI CV-QKD protocol can be seen as a special case of a typical one-way CV-QKD protocol~\cite{Grosshans2003,Weedbrook2004a}.

Then the secure key fraction can be estimated in accordance with the Devetak-Winter bound~\cite{Devetak2005,Pirandola2008}:
\begin{align}
r= \beta I(X_{\mathrm{A}},\:P_{\mathrm{A}}:X_{\mathrm{B}},\:P_{\mathrm{B}} )-\chi(X_{\mathrm{B}},\:P_{\mathrm{B}}:E),
\end{align}
where $0\leq \beta \leq1$ is the reconciliation efficiency (assumed to be ideal in the further numerical simulations), $I$ is the mutual information between Alice and Bob, $\chi(X_{\mathrm{B}} ,P_{\mathrm{B}}:E)=S(\widehat{\rho}_E)-S(\widehat{\rho}_E|X_{\mathrm{B}} ,P_{\mathrm{B}})$ is the Holevo bound, and $S(\widehat{\rho}_E)$ denotes the von Neumann entropy of quantum state $\widehat{\rho}_E$.

The upper bound $\chi(X_{\mathrm{B}},\:P_{\mathrm{B}}:\mathrm{A}_1,\mathrm{B}^{\prime}_1)$ is determined only using the corresponding covariance matrix. Supposing the system is under two independent entangling cloner attacks~\cite{Grosshans2003}, the covariance matrix takes the form of:
\begin{align}
\Xi&=\left(\begin{array}{cc}{V_{\mathrm{A}} I_2} & {\sqrt{(T(V_{\mathrm{A}}^2-1)\sigma_z )}} \\ {\sqrt{(T(V_{\mathrm{A}}^2-1)\sigma_z )}} & {[(V_{\mathrm{A}}-1)+1+T\xi^{\prime}]I_2}\end{array}\right),\\ T&=\frac{\eta_{\mathrm{A}}}{2} g^2,\\
\xi^{\prime}&=1+\frac{1}{\eta_{\mathrm{A}}}  [\eta_{\mathrm{B}} (\Xi_{\mathrm{B}}-1)+\eta_{\mathrm{A}} \Xi_{\mathrm{A}} ]+\frac{1}{\eta_{\mathrm{A}}} (\frac{\sqrt{2}}{g} \sqrt{V_{\mathrm{B}}-1}-\sqrt{\eta_{\mathrm{B}}}\sqrt{V_{\mathrm{B}}+1})^2,\\
\Xi_{\mathrm{A}}&=\frac{1-\eta_{\mathrm{A}}}{\eta_{\mathrm{A}}} +\xi_{\mathrm{A}},\:\Xi_{\mathrm{B}}=\frac{1-\eta_{\mathrm{B}}}{\eta_{\mathrm{B}}} +\xi_{\mathrm{B}},\\
\eta_{\mathrm{A}}&=10^{-\alpha L_{\mathrm{AC}}/10},\:\eta_{\mathrm{B}}=10^{-\alpha L_{\mathrm{BC}}/10},
\end{align}
where $\eta_{\mathrm{A}}\:(\eta_{\mathrm{B}})$ is a channel (Alice-Charlie or Bob-Charlie) transmittance, $\xi_{\mathrm{A}} (\xi_{\mathrm{B}})$ is the excess noise, $g$ is the offset factor, $I_2$ is the identity matrix, and $\sigma$ is the Pauli $z$-matrix.

To minimize the excess noise, the offset factor is set as
\begin{align}
g=\sqrt{\frac{2}{\eta_{\mathrm{B}}}}\sqrt{\frac{V_{\mathrm{B}}-1}{V_{\mathrm{B}}+1}}.
\end{align}

Then, the excess noise is expressed as:
\begin{align}
\xi^{\prime}=\xi_{\mathrm{A}}+\frac{1}{\eta_{\mathrm{A}}}[\eta_{\mathrm{B}} (\xi_{\mathrm{B}}-2)+2],
\end{align}
where the excess noise on Alice’s side (Bob’s side) $$\xi_{\mathrm{A(B)}}$$ contain the corresponding channel noise converted to SNU. Alice's and Bob's variances are considered equal $V_{\mathrm{A}} = V_{\mathrm{B}} = 40$ in the simulations. 

\section{Channel Noise Sources and their Mathematical Description}

Naturally, losses are inevitable when it comes to signal propagation in any medium, be it fiber-optical communication lines or free space. Regarding QKD, three effects are addressed in terms of noise primarily, which are the spontaneous Raman scattering (SpRS), the four-wave mixing (FWM) nonlinearity, and the linear channel crosstalk (LCXT). Now, we will briefly summarize their physical nature and corresponding mathematical description to then estimate the negative contribution they make to the performance of MDI CV-QKD system under consideration.  

\subsection{Spontaneous Raman Scattering}

Firstly, the main contributor to the overall channel losses in case QKD integrated with DWDM systems is the SpRS noise~\cite{Lin2021, Cia2019}. Its impact is minor for classical networks, though the contribution becomes substantial for joint QKD and DWDM systems~\cite{Lin2021, Cia2019}. The origin of the SpRS is different for the cases of co- and counter-propagation of signals. Thus, two sub-types are distinguished usually, that are forward (for co-propagating signals) and backward (for counter-propagating signals) SpRS noise. When speaking in the context of simultaneous QKD session and information transmission within single oprical fiber, the mathematical representation of their contribution is given by~\cite{Mlejnek2017, Eraerds2010}:

\begin{equation}
\label{ram_pw_f}
P_{\mathrm{ram,f}}=P_{\mathrm{out}}L\sum_{c=1}^{N_{\mathrm{ch}}}\rho (\lambda_{\mathrm{c}},\lambda _{\mathrm{q}})\Delta \lambda,
\end{equation}

and 

\begin{equation}
\label{ram_pw_b}
    P_{ram,b}=P_{out}\frac{\sinh(\xi L)}{\xi}\sum_{c=1}^{N_{ch}}\rho (\lambda _{c},\lambda _{q})\Delta \lambda, 
\end{equation}
where $P_{\mathrm{out}}$ denotes the output power for a single channel, $L$ is the length of the optical fiber,
$N_{ch}$ is the number of classical channels present in a DWDM system,
$\rho (\lambda_{\mathrm{c}},\lambda _{\mathrm{q}})$
describes the normalized scattering cross-section for the
wavelengths of classical ($\lambda_{\mathrm{c}}$) and quantum ($\lambda _{\mathrm{q}}$) channels, and $\Delta \lambda$ is the bandwidth of the quantum channel filtering system.

For the MDI CV-QKD realization considered here, both forward and backward SpRS occur for different system paths. A detailed description of the configuration will be provided in the following section.

The output power values appear in formulas instead of the input ones, so that to meet the the bit error rate (BER) requirements of a DWDM system directly. Thus, the value of $P_{\mathrm{out}}$ can be obtained as follows: 

\begin{equation}
\label{p_out}
P_{\mathrm{out}} \text{ (dBm)} = R_{\mathrm{x}} \text{ (dBm)} + IL \text{(dBm)},
\end{equation} 
where $R_{\mathrm{x}}$ is the sensitivity of a receiver and $IL$ denotes the insertion losses of the system. 

\subsection{Four-wave Mixing}

Next channel noise source to consider is FWM. It is a third-order nonlinear process, which sequence is creation of photons at new frequencies as a result of the interaction between the initial ones~\cite{Boyd}. These new frequencies might coincide with the one of the quantum channel~\cite{Lin2007}, thus contributing to the overall noise in the band of the quantum channel. 

To come up with the mathematical model for the FWM noise contribution consideration, let us consider three pump channels with the frequencies $f_{i}, f_{j}, \text{ and } f_{k}$. Then, the value of the resulting FWM noise peak power $P_{ijk}$ featuring the frequency $f_{i} + f_{j} - f_{k}$ can be expressed as~\cite{Mlejnek2017}:

\begin{equation}
\label{fwm_pwr}
    P_{ijk} = \eta \gamma^{2} D^{2} p^{2} e^{-\xi L} \frac{(1 - e^{ -\xi L})^{2}}{9 \xi^2} P_{s}P_{l}P_{h},
\end{equation} 
where the phase-matching efficiency for the FWM $\eta$ and parameter $\Delta \beta$ are defined as: 
\begin{equation}
    \eta = \frac{\xi^{2}}{\xi^{2} + \Delta \beta^{2}} \left[1 + \frac{4e^{- \xi L} \sin^{2}{(\Delta \beta L / 2)}}{(1 - e^{- \xi L})^{2}} \right],
\end{equation} 
and
\begin{equation}
    \Delta \beta = \frac{2 \pi \lambda^{2}}{c} \lvert f_{i} - f_{k} \rvert  \lvert f_{j} - f_{k} \rvert \cdot \left[D_{c} + \frac{d D_{c}}{d \lambda} \left(\frac{\lambda^{2}}{c} \right) \left(\lvert f_{i} - f_{k} \rvert + \lvert f_{j} - f_{k} \rvert \right) \right],
\end{equation}
correspondingly. In the above equations, $L$ is the transmission distance of the interacting light fields in the optical fiber, $D$ denotes the degeneracy factor ($D = 6$, $D = 3$), $P_{i(j,k)}$ and $f_{i(j,k)}$ are the input power and optical frequency of the interacting fields correspondingly, $\gamma$ stands for the third-order nonlinear coefficient, $\xi$ is the loss coefficient, $D_{c}$ and $d D_{c}/{d \lambda}$ are the dispersion coefficient of an optical fiber and its slope respectively with $\lambda$ standing for the wavelength of the FWM radiation.  

Finally, performing the summation of all the powers of the resulting FWM signals with frequencies coinciding with the frequency of the quantum channel, one obtains:

\begin{equation}
    P_{\mathrm{FWM}} = \sum P_{ijk},  f_{i} + f_{j} - f_{k} = f_{q}.
\end{equation} 

\subsection{Linear Channel Crosstalk}

It is due to the imperfections of the demultiplexers that any practically implemented DWDM system suffers LCXT losses~\cite{Hill1985}.

Since information signals are orders of magnitude more powerful than quantum ones, the insufficient isolation might cause considerable LCXT noise, that can be estimated in the following way:
\begin{equation}
    P_{\mathrm{LCXT}} = P_{\mathrm{out}} \text{ (dBm)} - \text{ISOL (dB)}.
\end{equation} 

Once the power values of the contributing to the overall channel noise effects are calculated, it is needed to recalculate them to a photon detection probability. To do so, the formula can be used:
\begin{equation}
\label{ram_prob}
    p_{\mathrm{ram,f(b)/FWM/LCXT}}=\frac{P_{\mathrm{ram,f(b)/FWM/LCXT}}}{hc/\lambda_{q}}\Delta t\eta_{D}\eta_{B},
\end{equation}
where $\eta_D$ denotes the detector efficiency, $\eta_{B}=10^{-0.1IL}$ is the transmission coefficient associated with the insertion losses  of the detection system, $h$ is the Planck constant, and $c$ is the light speed.

\section{MDI CV-QKD scheme and channel allocation}

Here, a possible realization scheme of MDI CV-QKD protocol will be addressed to analyze its potential for creating telecommunication optical transport networks integrated with DWDM systems. The notion of maximal achievable distances of MDI CV-QKD systems employed to characterize the latter denotes fiber length corresponding to the case, when the secure key generation rate is non-zero. 

\begin{figure}[ht]
\centering
\includegraphics[width=\textwidth]{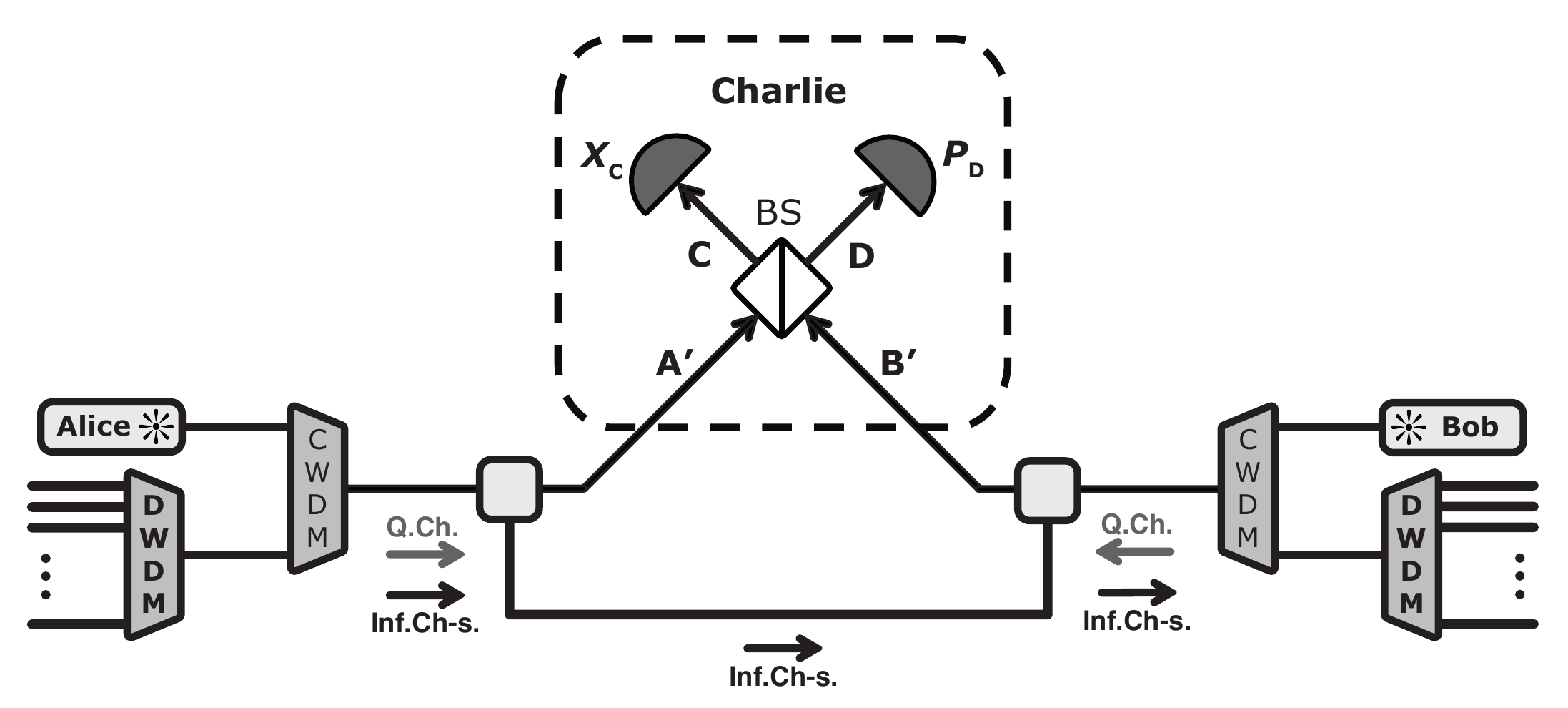}
\caption{Schematic illustration for the MDI CV-QKD protocol realization with DWDM: information is transmitted from Alice to Bob via DWDM-information channels; quantum channel is co-propagating with information channels in Alice's path, whereas counter-propagating with them in Bob's path and, thus, which means forward SpRS noise is induced in the Alice-Charlie path and backward SpRS - in the Bob-Charlie path}  
\label{fig:cv-mdi}
\end{figure}

The realization of MDI CV-QKD addressed in the work is shown in Figure~\ref{fig:cv-mdi}). 

Here, quantum signals are sent to the untrusted central relay to be homodyned there, whereas the information is transferred from Alice to Bob directly. It means, quantum and information signals are unidirectional in Alice's path (i.e., there is forward SpRS noise in the path) and counter-propagating through the Bob's path, so the SpRS noise features the backward type. 

The performance of the MDI CV-QKD realization was then numerically analyzed in terms of the possible channel allocation schemes and asymmetry coefficients between Alice's $L_{a}$ and Bob's $L_{b}$ paths $R_{\mathrm{asym}} = L_{a}/L_{b}$. 

Similarly to the work~\cite{Vorontsova2023}, four allocation schemes for the quantum channel located in C-band and O-band of the communication window were considered. The criterion and complete explanation for such a choice is provided in detail in the works~\cite{Vorontsova2022, Tarabrina2022}. The final choice of the configurations considered in the further numerical simulations is given in Table~\ref{tab_configs}. 

\begin{table}[ht]
\begin{center}
\caption{Description of the optimal configurations chosen for numerical simulations}
\label{tab_configs}
\begin{tabular}
{|c|c|c|c|}
\hline
Configuration &  \begin{tabular}[c]{@{}c@{}}Number of \\ channels\end{tabular}  & \begin{tabular}[c]{@{}c@{}}Quantum channel \\ wavelength, nm \end{tabular} \\
\hline 
Configuration 1 & 10 & 1536.61 \\
\hline
Configuration 2 & 10 & 1310 \\
\hline
Configuration 3 & 40 & 1537.40 \\
\hline
Configuration 4 & 40 & 1310 \\
\hline
\end{tabular}
\end{center}
\end{table}

The parameters of the DWDM system are the following: $\xi = 0.18$~dB/km, $\Delta \lambda = 15$~GHz, $N_{ch} = 10 \text{ or } 40$, $R_{x} = -32$~dBm and IL~=~8~dB. 

As for the asymmetry coefficient $R_{\mathrm{asym}}$, three different relations are address here: a symmetric (i.e., $L_{a}/L_{b} = 1$) and two asymmetric realizations ($L_{a}/L_{b} = 3/2$ and $L_{a}/L_{b} = 2/1$). 

\section{Results and discussion}

\begin{figure}[ht]
\centering
\includegraphics[width=0.86\textwidth]{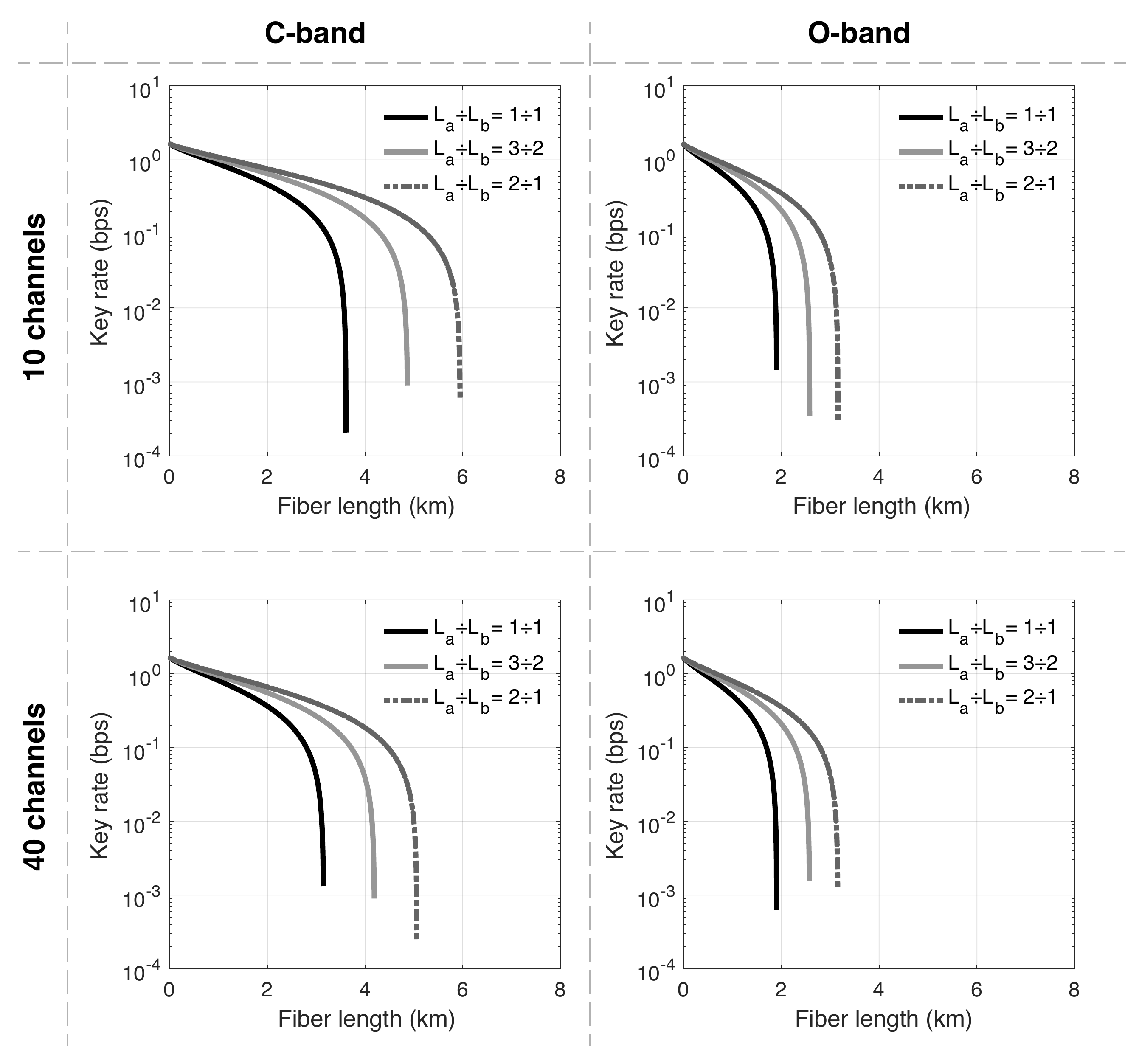}
\caption{The dependence of the secure key generation rate on the optical fiber length for MDI CV-QKD corresponding to the scheme in Figure~\ref{fig:cv-mdi}.}  
\label{fig:cv-mdi_res}
\end{figure}

Using the mathematical models for the MDI CV-QKD secure key generation rate, SpRS, FWM, and LCXT noises, the realization of the system depicted in Figure~\ref{fig:cv-mdi} was numerically simulated. The results are presented in Figure~\ref{fig:cv-mdi_res}. 

It is a known fact that for MDI CV-QKD systems the secure key generation rate decreases dramatically as the system approaches its symmetry ($L_a = L_b$), with the best result corresponding to the situation when one of the paths equals zero~\cite{Li2014}. 

The results obtained confirm the stated conclusion, as the maximal achievable distance increases together with the system's paths asymmetry. 

Interestingly, for the case of MDI CV-QKD, configurations for which the quantum channel is located in the C-band appeared to be more efficient in terms of their maximal achievable distances. For most cases, configurations with O-band-located quantum channel are more beneficial (MDI QKD as well, see~\cite{Vorontsova2022}), as FWM noise does not contribute to the overall losses. Though the overall contribution of channel noises here is less for such cases too, fiber attenuation for the quantum channel wavelength of 1310~nm surpasses them significantly. 
It can be seen also that the maximum achievable distances do not exceed 6~km, thus, the secure key distribution over long distances is not possible here. Still, such realizations can be utilized for short-distance communication. 

Regarding the number of information channels (10 and 40 in the work), for a larger number of information channels, a decrease in the maximal achievable distances takes place. This is a natural observation, as the more information channels are there in the system, the greater the value of the overall channel losses. The decrease is substantial for C-band configurations, whereas is quite small in case of O-band. 

\section{Conclusion}
\label{sec:conc}
In this work, the MDI CV-QKD protocol was addressed. A theoretical research and numerical simulation of the noise influence caused by SpRS, FWM, and LCXT effects on the performance of the MDI CV-QKD system performance was analyzed for a proposed practical realization scheme, in terms of channel allocation and the paths' asymmetry coefficient. Increasing the number of information channels naturally leads to a decrease in the maximal achievable distances. In addition, the allocation of a wavelength of 1310~nm for a quantum channel results in the shortening of maximal distance values for MDI CV-QKD, regardless of the fact the overall channel noise is less for such configurations. The superior contribution comes from the fiber attenuation, which is larger for the wavelength of 1310~nm. It was confirmed that the more asymmetric the paths for the MDI CV-QKD scheme are, the more efficient the systems is. Moreover, MDI CV-QKD realizations feature significantly shorter maximal achievable distances, that do not exceed several kilometers and can be utilized for short-distance information transmission only. The results obtained can be used in terms of practical implementation of MDI CV-QKD systems, so that to obtain optimal results.

\section*{Acknowledgements}
The work was done by Leading Research Center "National Center for Quantum Internet" of ITMO University by the order of JSCo Russian Railways.

\end{document}